\documentclass[aps,english,prb,reprint,superscriptaddress, citeautoscript,cite]{revtex4-1}

\usepackage{graphicx}
\usepackage{amssymb}
\usepackage{babel}
\usepackage{color}
\usepackage{multirow}
\usepackage{float}

\begin{document}

\title{Electronic and magnetic properties of the monolayer RuCl$_3$: 	A first-principles and Monte Carlo study}

\author{S. Sarikurt}
\affiliation{Dokuz Eyl\"{u}l University, Faculty of Science, Physics Department, T{\i}naztepe Campus, 35390 İzmir, Turkey}
\author{Y. Kadioglu}
\affiliation{Department of Physics, Adnan Menderes University, Aydin 09010, Turkey}
\author{F. Ersan}
\affiliation{Department of Physics, Adnan Menderes University, Aydin 09010, Turkey}
\author{E. Vatansever}
\affiliation{Dokuz Eyl\"{u}l University, Faculty of Science, Physics Department, T{\i}naztepe Campus, 35390 İzmir, Turkey}
\author{O.\"{U}zengi Akt\"{u}rk }
\affiliation{Department of Electrical and Electronic Engineering, Adnan Menderes University,
09100 Ayd{\i}n, Turkey}
\affiliation{Nanotechnology Application and Research Center, Adnan Menderes University, Aydin 09010, Turkey}
\author{Y.Y\"{u}ksel}
\affiliation{Dokuz Eyl\"{u}l University, Faculty of Science, Physics Department, T{\i}naztepe Campus, 35390 İzmir, Turkey}
\author{U. Ak{\i}nc{\i}}\email{umit.	akinci@deu.edu.tr}
\affiliation{Dokuz Eyl\"{u}l University, Faculty of Science, Physics Department, T{\i}naztepe Campus, 35390 İzmir, Turkey}
\author{E. Akt\"{u}rk}\email{ethem.akturk@adu.edu.tr}
\affiliation{Department of Physics, Adnan Menderes University, Aydin 09010, Turkey}
\affiliation{Nanotechnology Application and Research Center, Adnan Menderes University, Aydin 09010, Turkey}
\date{\today}



\begin{abstract}
\noindent\normalsize{Recent experiments revealed that monolayer $\alpha$-RuCl$_3$ can be obtain by chemical exfoliation method and exfoliation or restacking of nanosheets can manipulate the magnetic properties of the materials. In this present paper, the electronic and magnetic properties of $\alpha$-RuCl$_3$ monolayer are investigated by combining first-principles calculations and Monte Carlo simulations. From first-principles calculations, we found that the spin configuration FM corresponds to the ground state for $\alpha$-RuCl$_3$, however, the other excited zigzag oriented spin configuration has energy of  5 meV/atom higher than the ground state. Energy band gap has been obtained as $3$ meV using PBE functionals. When spin-orbit coupling effect is taken into account, corresponding energy gap is determined to be as $57$ meV. We also investigate the effect of Hubbard U energy terms on the electronic band structure of $\alpha$-RuCl$_3$ monolayer and revealed band gap increases approximately linear with increasing U value. Moreover, spin-spin coupling terms ($J_1$, $J_2$, $J_3$) have been obtained using first principles calculations. By benefiting from these terms, Monte Carlo simulations with single site update Metropolis algorithm have been implemented to elucidate magnetic properties of the considered system. Thermal variations of magnetization, susceptibility and also specific heat curves indicate that monolayer $\alpha$-RuCl$_3$ exhibits a phase transition between ordered and disordered phases at the Curie temperature $14.21$ K. We believe that this study can be utilized to improve two-dimensional magnet materials.}
\end{abstract}

\maketitle
\section{INTRODUCTION}
Since the discovery of the graphene~\cite{Novoselov2004}, two-dimensional (2D) materials are highly attractive for potential applications in electronic, spintronic and magnetoelectronic devices~\cite{Butler_2013, Xu_2013, Chhowalla_2013, Bhimanapati_2015, Tan_2017}. The former studies indicate that the low-dimensional materials such as graphene~\cite{Novoselov2004}, silicene~\cite{Cahangirov2009}, boron nitride~\cite{ZhangGuo2012}, zinc-oxide~\cite{wang2004zinc}, phosphorene~\cite{liu2014phosphorene}, arsenene~\cite{kamal2015arsenene}, antimonene~\cite{olcay} and bismuthene~\cite{akturk} exhibit a variety of new electronic properties from their bulk phases. In addition to these, investigation of the properties of one- and two-dimensional transition metal chalcogen (TMC) compounds (in the forms of MX$_i$ with $i=1,2,3$ and X=O, S, Se, Te) has recently received considerable attention because of the fact that the TMCs are generally found to be in three-dimensional (3D) layered structure in which the interlayer is of van der Waals (vdW) type.  One of the remarkable properties of these materials which makes them important in nanotechnology is that these TMC compounds have direct energy band gap between 1-2 eV whereas they have indirect energy band gap in 3D phase. The common feature of these aforementioned structures that they are all non-magnetic semiconductor materials. Magnetization can be achieved by doping the material with foreign atoms, generating defects in the material, applying external force or tensile strain~\cite{MaDai2012, Zhou2012}. In recent studies, researchers indicate that it is possible to find 2D materials that have robust intrinsic magnetism. For instance, 2D MnX$_2$ (X=O, S, Se) materials exhibit ferromagnetism ($3\mu_B$) with Curie temperatures ($T_C$) $140\ K$, $225\ K$ and $250\ K$, respectively~\cite{Kan2013, Kan2014}.  Another group of 3D materials containing transition metals and weak vdW interaction between the layers are metal halides called as MX$_3$ (M = Ti, V, Cr, Fe, Mo, Ru, Rh, Ir and X = Cl, Br, I). These three-dimensional materials were synthesized years ago~\cite{Bengel1995, Hillebrecht1997}. The spacing between the sheets of these materials layered with the ABC packing ranges from $3$ to $3.50$ {\AA} and is suitable to reduce the dimension from 3D to 2D~\cite{Zhou2016}. Hence, in recent years, experimental studies have also been conducted to produce 2D layered forms of metal halogens, and it has been proved that this kind of materials can be experimentally synthesized~\cite{miro2014atlas}.

There also exists a rapidly increasing theoretical research activity focused on this area~\cite{McGuire2017crystal}. For instance,
Liu et al.~\cite{Liu2016} investigate the ferromagnetism in monolayer Cr-trihalides (CrX$_3$ (X=Cl, Br, I)) using density functional
theory (DFT) combined with the self-consistently determined Hubbard U approach (DFT+U$_{scf}$) and Monte Carlo (MC) simulations that
based on Ising model. CrBr$_3$ was obtained as the first ferromagnetic (FM) semiconductor in 1960~\cite{Tsubokawa}. Liu and coworkers
reported that the FM state is always the ground state for CrX$_3$ (X = Cl, Br, I) monolayer structures with a net magnetic
moment of $6\mu_B$ in each unit cell. Also, they point out that Cr ions give the main contribution to magnetic moment and the
neighboring X (X = Cl, Br, I)  ions are spin polarized as antiferromagnetic (AFM). They calculate the exchange-correlation
energy per unit cell ($E_{ex}=E_{AF}-E_{FM}$) for CrCl$_3$, CrBr$_3$, CrI$_3$ as $34$, $44$ and $55$ meV, respectively.
These results represent that ferromagnetic (FM) state of each CrX$_3$(X=Cl, Br, I) monolayers has the lowest energy and
that corresponds to the ground state. So, these monolayer chromium trihalides are reported as a series of stable 2D
intrinsic FM semiconductors. CrCl$_3$, CrBr$_3$, CrI$_3$ monolayers are semiconductors with energy gap $2.28$, $1.76$ and $1.09$ eV, respectively.
Only the interaction between first nearest neighbors was taken into account in the exchange interaction of these monolayer
structures and exchange interaction parameters of CrCl$_3$, CrBr$_3$, CrI$_3$ obtained as $0.63, 0.81$ and $1.02$ meV, respectively.
The calculated Curie temperature values for CrCl$_3$, CrBr$_3$, CrI$_3$ are $66$, $86$ and $10\ K$. The Curie temperature values
obtained for CrI$_3$ by different research groups are as follows; $61\ K$~\cite{McGuire2015} and $75\ K$~\cite{HWang2016}.
For CrX$_3$ (X=F, Cl, Br, I), Zhang et al. have obtained respective Curie temperature such as $41, 49, 73$ and $95\ K$~\cite{Zhang2015}.
They reported the energy band gap values as $4.68, 3.44, 2.54$ and $1.53$ eV for CrF$_3$ , CrCl$_3$, CrBr$_3$, CrI$_3$. According to their
calculation results, these 2D Cr-trihalides are half semiconductors with indirect band gaps. For magnetization calculations, they considered the nearest-, next-nearest- and the next-next-nearest neighbor exchange-coupling parameters different from Liu et al.~\cite{Liu2016}.
The nearest- and next-nearest-neighbor exchange-coupling parameters are indicated as FM, the next-next-nearest neighbor exchange-coupling parameter as AF for these single layer Cr-trihalides.

He et al.~\cite{HeMa2016} obtained the stability, electronic and magnetic properties of VCl$_3$ and VI$_3$ monolayers using DFT+U$_{scf}$ approach
together with the MC simulations. Their results represent that VCl$_3$ and VI$_3$ monolayers are dynamically stable
and also these monolayers exhibit an intrinsic ferromagnetism and half-metallicity. They found that the total
magnetic moment per unit cell is $4\mu_B$. According to Heisenberg Hamiltonian, they calculated the nearest-, next-nearest-
and the next-next-nearest neighbor exchange-coupling parameters $J_1$, $J_2$, $J_3$ as $2.227, 0.144$ and $0.02$ meV for VCl$_3$
and $2.754, -0.019$ and $0.110$ meV for VI$_3$, respectively. So, the magnetic interaction of VI$_3$ single layer structure is
greater than VCl$_3$. They reported the respective Curie temperature of VCl$_3$ and VI$_3$ layers as $80\ K$ and $98\ K$. They also
investigated the effect of the electron and hole doping on ferromagnetism and Curie temperature. The carrier doping leads to an
enhancement in ferromagnetism of VCl$_3$ and VI$_3$ monolayers. And the Curie temperature of doped VCl$_3$ and VI$_3$ increase
with increasing the carrier concentration.
\newline
\indent The Ruthenium Chloride (RuCl$_3$) structure known for its interesting catalytic~\cite{Fieser1990, Shim1996} and
photochemical~\cite{Matsuoka1991} properties has recently attracted much attention with its magnetic properties. Theoretical
and experimental analysis of magnetic and thermodynamic properties for $\alpha$-RuCl$_3$ have been examined in many
studies~\cite{sears2015magnetic, johnson2015monoclinic, banerjee2016proximate, banerjee2017neutron, Aoyama2017, wolter2017field, janssen2017magn}.
And besides, the interplay between spin-orbit coupling and electronic  correlations have been performed~\cite{plumb2014alpha, kim2015kitaev}. 
For the materials with strong spin-orbit coupling (SOC), there exist bond dependent interactions ~\cite{sears2015magnetic, janssen2016, janssen2017}. Magnetic properties of these kind of materials  can be modeled by the Kitaev-Heisenberg (KH) spin liquid model ~\cite{Kitaev2006} which is defined on a honeycomb lattice. Using this model, field induced magnetization processes in $\alpha$-RuCl$_3$ and $A_2IrO_3$(A=Na,Li)  have been investigated ~\cite{johnson2015monoclinic, janssen2016, janssen2017}. For $\alpha$-RuCl$_3$, KH model has been recently proposed by Banerjee et al. ~\cite{banerjee2016proximate}. In particular, Price and Perkins ~\cite{Price2012, Price2013} have performed Monte Carlo simulations based on standard Metropolis algorithm for the investigation of classical KH model. They found that there exists Berezinskii-Kosterlitz-Thouless (BKT) critical behavior before the order-disorder transition takes place.

Most recently, Huang et al.~\cite{Huang2017} have calculated the electronic and magnetic properties of RuI$_3$ monolayer by using spin-polarized DFT and MC simulation. They have obtained the nearest-neighbor exchange-coupling parameter as $82\ meV$ and the Curie temperature approximately $T_C=360\ K$, which is higher than most of the 2D FM nanomaterials studied heretofore. They have also investigated the in-plane strain effect on magnetic exchange and the global band gap of RuI$_3$ monolayer. Based on the results,
they have pointed out that RuI$_3$ monolayer is an intrinsic FM quantum anomalous Hall insulator. Besides these calculations, they
have also analyzed the FM configurations of RuCl$_3$ and RuBr$_3$ monolayers. Weber et al. chemically exfoliated the monolayer $\alpha$-RuCl$_3$, but they did not clarify the magnetism of a suspended monolayer on SiO$_2$/Si substrate~\cite{weber2016magnetic}. In another study, Ziatdinov et al. mentioned in their study that magnetic ground state can occurs in the geometry of the ligand sublattice in thin films of RuCl$_3$~\cite{ziatdinov2016atomic}. To the best of our knowledge, the investigation of the possibility of magnetic ground state for $\alpha$-RuCl$_3$ monolayer have not been yet thoroughly studied. In this paper, we systematically investigate the electronic and magnetic properties of $\alpha$-RuCl$_3$ monolayer using DFT and MC simulations.  The magnetic exchange coupling constants have been obtained from DFT calculations. The Curie temperature was calculated by using these exchange-coupling parameters in MC simulations based on the Heisenberg model. The calculated Curie temperature of $\alpha$-RuCl$_3$ monolayer is $T_C = 14.21\ K$. Our results indicate that monolayer $\alpha$-RuCl$_3$ is stable two-dimensional intrinsic ferromagnetic semiconductor.

\section{Computational Methodology}\label{comp}
The calculations for electronic and magnetic properties of $\alpha$-RuCl$_3$ have been carried out using DFT with the Projector
Augmented Wave (PAW) method~\cite{Blochl1994, Kresse1999ultrasoft} as implemented in the Vienna ab initio Simulation
Package (VASP)~\cite{Kresse1993, Kresse1994, Kresse1996a, Kresse1996b}. We have used Generalized Gradient
Approximation (GGA) in the Perdew-Burke and Ernzerhof (PBE)~\cite{Perdew1992atoms, Perdew1996generalized} form for
exchange correlation potential. Furthermore, we investigate SOC effects on the electronic
structure of $\alpha$-RuCl$_3$.  An energy cutoff of $400\ eV$ was used with Gamma-centered Monkhorst-Pack~\cite{MP1976}
special k-point grids of 16$\times$16$\times$1 and 16$\times$8$\times$1 for bulk $\alpha$-RuCl$_3$ and monolayer, respectively.
The convergence criterion for total energy is assumed as $10^{-5}\ eV$ and the maximum force of 0.002 eV/\AA{} was
allowed on each atom. We consider a vacuum layer of about $20$ {\AA} thick along the z-axis to eliminate the
interaction between neighboring layers. We obtain the phonon dispersion using the finite
displacement approach as defined in PHONOPY code\cite{phonopy} for 2$\times$2$\times$1 supercell and a
displacement of $0.01$ {\AA} from the equilibrium atomic positions. To investigate the thermal stability of the optimized
$\alpha$-RuCl$_3$ monolayer, we performed \textit{ab initio} molecular dynamics (MD) calculations. Verlet algorithm and Nos\'{e} thermostat was used for this examination. We also check the structure with
Quantum Espresso (QE) software\cite{qe} by using Vanderbilt ultrasoft type pseudopotential. We choose again
GGA and PBE parametrisation.

To determine the magnetic structure of $\alpha$-RuCl$_3$, the nearest-, next-nearest- and next-next-nearest-neighbors exchange-coupling parameters (J$_1$, J$_2$ and J$_3$, respectively), we fit the total energies obtained from DFT calculations for different magnetic configurations to the Heisenberg Spin Hamiltonian:

\begin{equation}
H_{spin}=-J_1\sum_{ij}  S_i S_j - J_2\sum_{kl}  S_k S_l - J_3 \sum_{mn}  S_m S_n
\end{equation}
where $S_i$ is the spin at the Ru site i and (i, j), (k, l) and (m, n) stand for the nearest,
next-nearest and next-next-nearest Ru atoms, respectively. The in-plane ($E[100]-E[010]$) and out-of plane ($E[100]-E[001]$) magnetic anisotropy energies (MAEs)  are obtained as $61\ \mu eV$ and $-18.88\ meV$, respectively. This negative out-of plane MAE denotes that $\alpha$-RuCl$_3$ monolayer has in-plane magnetic anisotropy (in-plane easy axis of magnetization).

By mapping the DFT energies to the Heisenberg Spin Hamiltonian, J$_1$, J$_2$ and J$_3$ can be calculated from following equations~\cite{Sivadas2015}:
\begin{equation}
E_{FM/Neel}=E_0 - (\pm 3J_1 + 6J_2 \pm 3J_3)S^2
\end{equation}
and
\begin{equation}
E_{Zigzag/Stripy}=E_0 - (\pm J_1 - 2J_2 \mp 3J_3)S^2
\end{equation}

The J$_1$, J$_2$ and J$_3$  values are found to be $10.69$ meV, $-1.26$ meV and $2.54$ meV, respectively. The first and the third neighboring exchange parameters are FM while the second neighboring exchange parameter is AFM. The Curie temperature was calculated by using these exchange-coupling parameters in MC simulations based on the Heisenberg model.

\section{Model and DFT Simulation Details}\label{model}
\begin{figure}[h]
\includegraphics[scale=0.5]{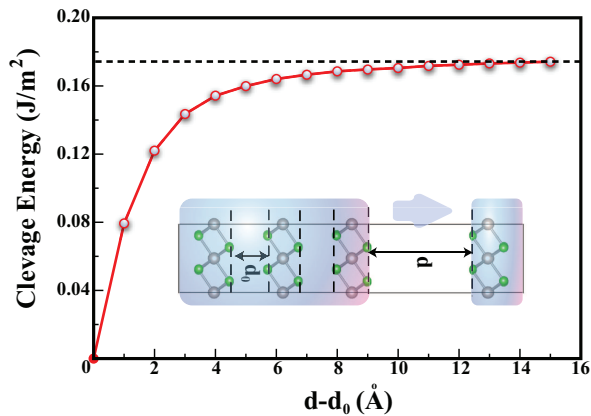}
\caption{(Color online) Cleavage energy as a function of the separation between two fractured parts. The fracture distance defined as $d$ and the equilibrium interlayer distance of ruthenium trihalides as $d_0$. Inside the graph: Side view of bulk $\alpha$-RuCl$_3$ used to simulate the exfoliation procedure}
\label{fig1}
\end{figure}

\begin{figure}[h]
\includegraphics[scale=0.45]{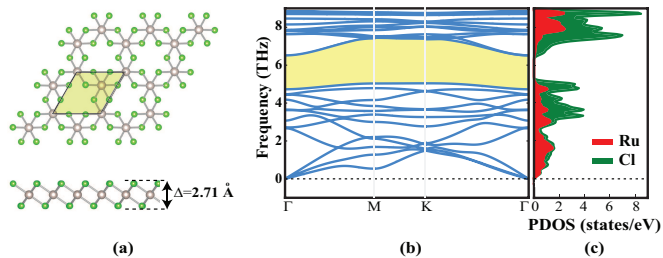}
\caption{(Color online) (a) Top and side views of the optimized RuCl$_3$ monolayer (b) Phonon
frequencies along the main symmetry directions in 2D Brillouin zone}
\label{fig2}
\end{figure}

$\alpha$-RuCl$_3$ has ABC type stacking in layered and has \textit{C2/m} space group for it's bulk form as illustrated in Figure~\ref{fig1}. As seen in Figure~\ref{fig1}, layered $\alpha$-RuCl$_3$ consist of Cl-Ru-Cl sandwich layers and each Ru atom is located at an octahedral site between the Cl atom layers. Therefore while the Cl atoms form a hexagonal pattern, the metal atoms have a honeycomb lattice. Before studying the monolayer form of $\alpha$-RuCl$_3$, we first tested the possibility of exfoliation process to obtain monolayer from its bulk form. First, we created bulk $\alpha$-RuCl$_3$ with a=$5.99$, b=$10.37$ and c=$6.05$ \AA{} lattice constants as illustrated in Figure~\ref{fig1} (suitable with literature~\cite{weber2016magnetic, ziatdinov2016atomic, cao2016low}) and we implemented a fracture in the bulk after four periodic layer and we systematically increase this fracture distance, at the end we calculate the corresponding cleavage energy (CE) (Figure~\ref{fig1}). This method is very effective and has been widely confirmed~\cite{Zhang2015, McGuire2015, HeMa2016, Liu2016}. We add separately two van der Waals (vdW) correction terms in the calculations. One of them is the most common used DFT-D2\cite{DFT-D2} correction term and the other one is Tkatchenko and Scheffler (TS) method,\cite{TS} which is formally identical to that of DFT-D2 method. However, in TS method the dispersion coefficients and damping function are charge-density dependent, differently from DFT-D2 method. Obtained CE is $0.174$ J/m$^2$ for DFT-D2 and $0.238$ J/m$^2$ for TS correction terms and these energy values are approximately half of the CE of graphite (0.39$\pm$0.02 J/m$^2$),\cite{graphiteCE} though this energy value can be increase with used vdW correction terms, but we suppose that CE value of $\alpha$-RuCl$_3$ will be comparable with graphite. And also this result is comparable with those of the other MX$_3$ (M=Cr, V, Ti ; X=Cl, Br, I) materials~\cite{Liu2016,HeMa2016}. Consequently, according to this result $\alpha$-RuCl$_3$ monolayer can be obtained by the exfoliation process from its bulk phase whereas realized recently by Weber et  al.~\cite{weber2016magnetic} Secondly, we obtain the monolayer $\alpha$-RuCl$_3$ (see Figure~\ref{fig2}(a)), and to confirm the dynamical stability of $\alpha$-RuCl$_3$ we calculate the phonon frequencies (see Figure~\ref{fig2}(b)) along the main symmetry directions in 2D Brillouin zone (BZ) by using the PHONOPY program, which is based on the finite-displacement method as implemented in VASP.\cite{phonopy} Our calculations show that there is not any negative frequencies in the BZ, which proved the dynamical stability at T$\sim$0 K temperature.

In addition, if we see phonon dispersion in more detail, we can see a local minima in the Brillouin zone (at M high symmetry point) for out of plane acoustical branch (ZA). This kind of minima in the phonon bands are associated with Kohn anomalies and cause to discontinuities in $\delta \Omega / \delta k$. In fact, this minima occurs at the high symmetry k-points in the BZ due to singularity of the dielectric function. This situation has been observed in VCl$_3$, VI$_3$, NiCl$_3$, and some MX$_2$ monolayers~\cite{HeMa2016,nicl3,Salim_Hoca}. In addition, we checked the thermal stability by ab initio MD calculation for 2 ps at $300$ K and $700$ K, at the end of MD simulations there is not observed any structural deformation, except of small thermal fluctuations.

\begin{figure}[h]
\includegraphics[scale=0.4]{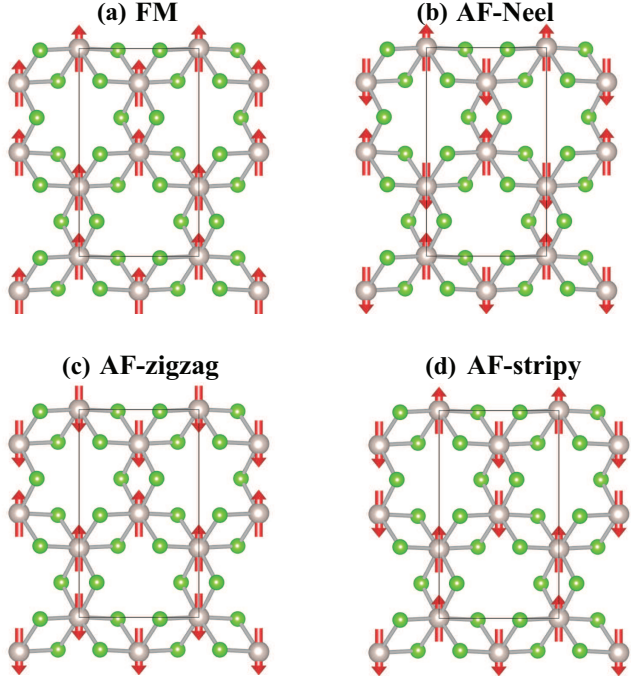}
\caption{(Color online) Different spin configurations of RuCl$_3$ monolayer (a) FM ordered (b) AF-N\'{e}el ordered (c) AF-Zigzag ordered and (d) AF-stripy ordered.}
\label{fig3}
\end{figure}

\begin{figure}[h]
\includegraphics[scale=0.45]{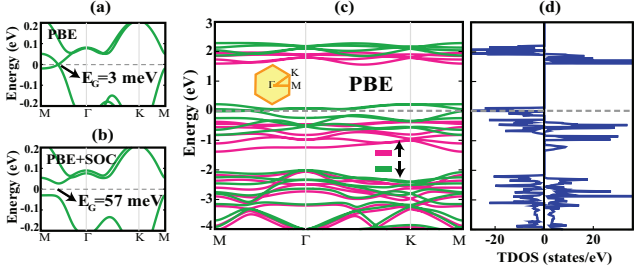}
\caption{(Color online) Electronic band structure of monolayer RuCl$_3$ (a) without SOC (b) with SOC.  (c)
Band structures calculated using PBE functionals and (d) total electronic density of states for monolayer RuCl$_3$}
\label{fig4}
\end{figure}

\begin{figure}[h]
\includegraphics[scale=0.45]{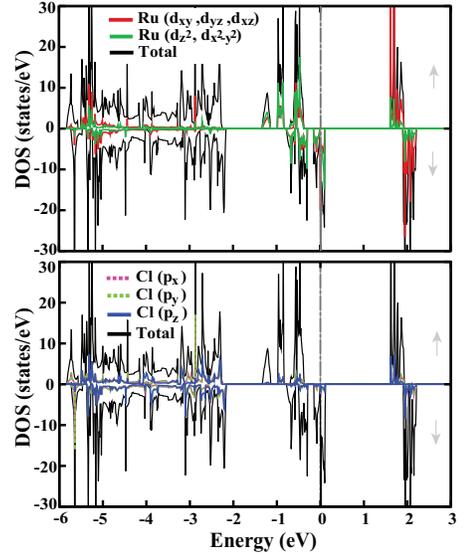}
\caption{(Color online) The partial and total density of states for the ground states of (a) Ru and (b) Cl atoms of monolayer RuCl$_3$}
\label{fig5}
\end{figure}

To determine the favorable magnetic ground state structure of $\alpha$-RuCl$_3$, we considered four types of spin configurations (FM, AFM-N\'{e}el, AFM-Zigzag and AFM-Stripy) for Ruthenium atoms as shown in Figure~\ref{fig3}. We found FM state is the most stable magnetic configuration and RuCl$_3$ has 4$\mu_B$ total magnetic moment  (per square unit cell). The other excited configurations, AFM-N\'{e}el, AFM-Zigzag and AFM-Stripy have energies higher than the ground state. Relative energy values for other spin oriented systems can be find in S.I Table1. In this work the electronic band structure and total electronic density of states of FM-RuCl$_3$ are given in Figure~\ref{fig4}. While spin-up channel has large band gap ($1.93$ eV), spin down channel has almost matching bands between the M and $\Gamma$ high symmetry points. There is just $3$ meV band gap between the valance band maximum (VBM) and conduction band minimum (CBM) for PBE calculation, however these bands are seperated from each other and 57 meV band gap occurs with implemented SOC effect in the calculation. Partial density of states (PDOS) are given in Figure~\ref{fig5} to show the contribution of orbital around the Fermi level. As seen in Figure~\ref{fig5} opposite to graphene DOS, p$_z$ does not contribute significantly at Fermi level for spin-down channel, dominant contribution comes from Ru d-orbital. Two fold e$_g$ (d$z^2$ and d$x^2$-d$y^2$) orbitals and three fold t$_{2g}$ (d$_{xy}$, d$_{yz}$ and d$_{xz}$) orbitals give approximately equal contribution to valance bands for both spin-up and -down channels. Cl-p orbitals are dominantly shown at lower energy values than -2 eV. We also checked and confirmed the electronic structure of RuCl$_3$ by QE and found similar band structure with VASP calculation. Kim et al. found that monolayer $\alpha$-RuCl$_3$ shows metallic character by LDA calculation, however it can turn to semiconductor by adding SOC and Hubbard parameter (U) to classical LDA calculation~\cite{kim2015kitaev}, and their general trend of band structure is compatible with our band structure. But, unfortunately there is a discrepancy between our band structure and Huang et al. results for RuCl$_3$~\cite{Huang2017}. They found metal for RuCl$_3$ both of PBE and PBE+SOC calculation and found approximately lineer Dirac point at \textbf{K} high symmetry point, while  Kim et al. and  Li et al. found band matching between the \textbf{$\Gamma$}-\textbf{M} symmetry points ~\cite{kim2015kitaev,Li_2017}. Our result is in fair agreement 	with Kim et al. and  Li et al. . As mentioned literature, the electronic and magnetic properties of bulk $\alpha$-RuCl$_3$ can be described properly only when a proper Hubbard U term and SOC term is added. Therefore, we investigate the magnetic ground state and electronic properties of single layer RuCl$_3$ by varying U energy terms. For these calculations we reoptimized the structures to obtain ground state energies for added terms. Relative energies can be find in S.I Figure I, according to our results FM configuration energetically is the most favorable configuration, but for high U values ($2.5$ eV, $3.0$ eV) Zigzag and FM configurations have similar energies, and Neel configurations is the second favorable configuration. Our results also show that band gap of monolayer RuCl$_3$ has nearly directly proportional with increasing U term, and for U=$1.0$ and $1.5$ eV band gaps of FM and Zigzag configurated RuCl$_3$ are approximately $0.27$ eV which is similar with experimental result.~\cite{ziatdinov2016atomic}

\section{Monte Carlo Simulation Details}
In order to understand and clarify magnetic properties of the considered $\alpha$-RuCl$_3$ monolayer, we use Monte Carlo simulation method with single-site update Metropolis algorithm~\cite{Newman1999MC, Binder1979MC}. The system is located on a $L\times L$ honeycomb lattice, where $L$ is the linear size of the system, and throughout the study we have fixed as $L=120$. There is a Ru atom in each lattice point in the system. Periodic boundary conditions are applied in all directions. The simulation procedure can be briefly summarized as follows. We note that numerical data are generated over $100$ independent sample realizations. In each sample realization, the simulation starts from $T=30.01\ K$ using random initial configurations. After, it is slowly cooled down until the temperature reaches $T=10^{-2}\ K$ with small temperature steps $\Delta T =0.2\ K$. Spin configurations are generated by selecting the site sequentially through the honeycomb lattice. For each temperature, typically first $10^4$ Monte Carlo step per site (MCSS) are discarded for thermalization process. Next, $9\times 10^4$ MCSS are used to collect and to determine the temperature dependencies of the physical quantities.

Instantaneous  magnetization components of the system can be defined as follows:
\begin{equation}
m_\alpha=\frac{1}{N}\sum_{i=1}^{N}S_{i}^{\alpha},
\end{equation}\label{Eq1MC}
here $N$ is the total number of the spins in the RuCl$_3$ monolayer and $\alpha=x, y$ and $z$. By means of Eq. \ref{Eq1MC}, the magnetization of the system can be given as follows:
\begin{equation}
 M=\sqrt{m_{x}^2+m_{y}^2+m_{z}^2}.
\end{equation}

In order to determine Curie point of the considered system, we use the thermal
variations of the magnetic susceptibility,
\begin{equation}
\chi=N\left(\langle M^2 \rangle-\langle M \rangle^2\right)/k_BT,
\end{equation}

\noindent and specific heat  curves,
\begin{equation}
C=N\left(\langle E^2 \rangle-\langle E \rangle^2\right)/k_BT^2.
\end{equation}

\begin{figure}[h]
\includegraphics[scale=0.3]{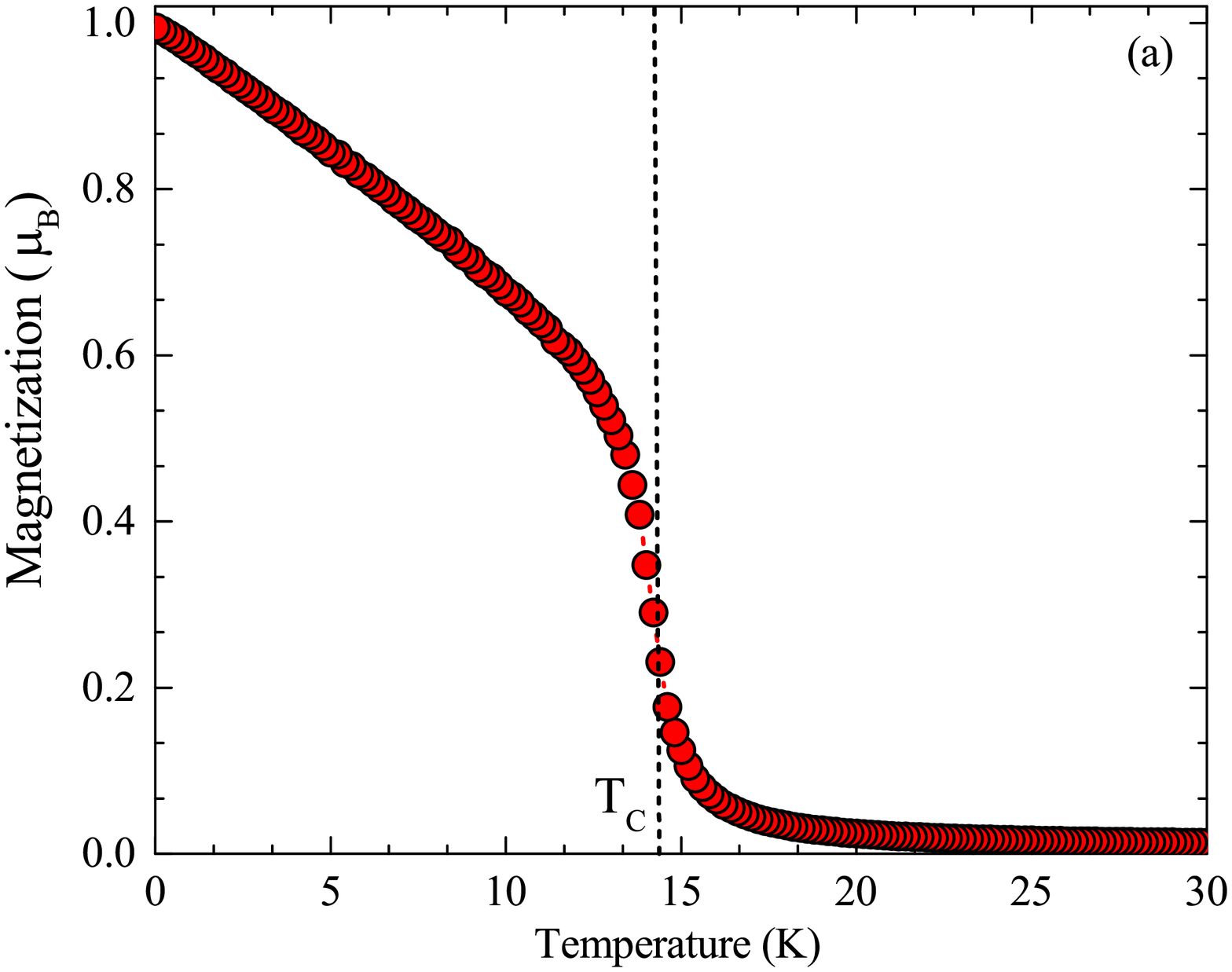}
\includegraphics[scale=0.3]{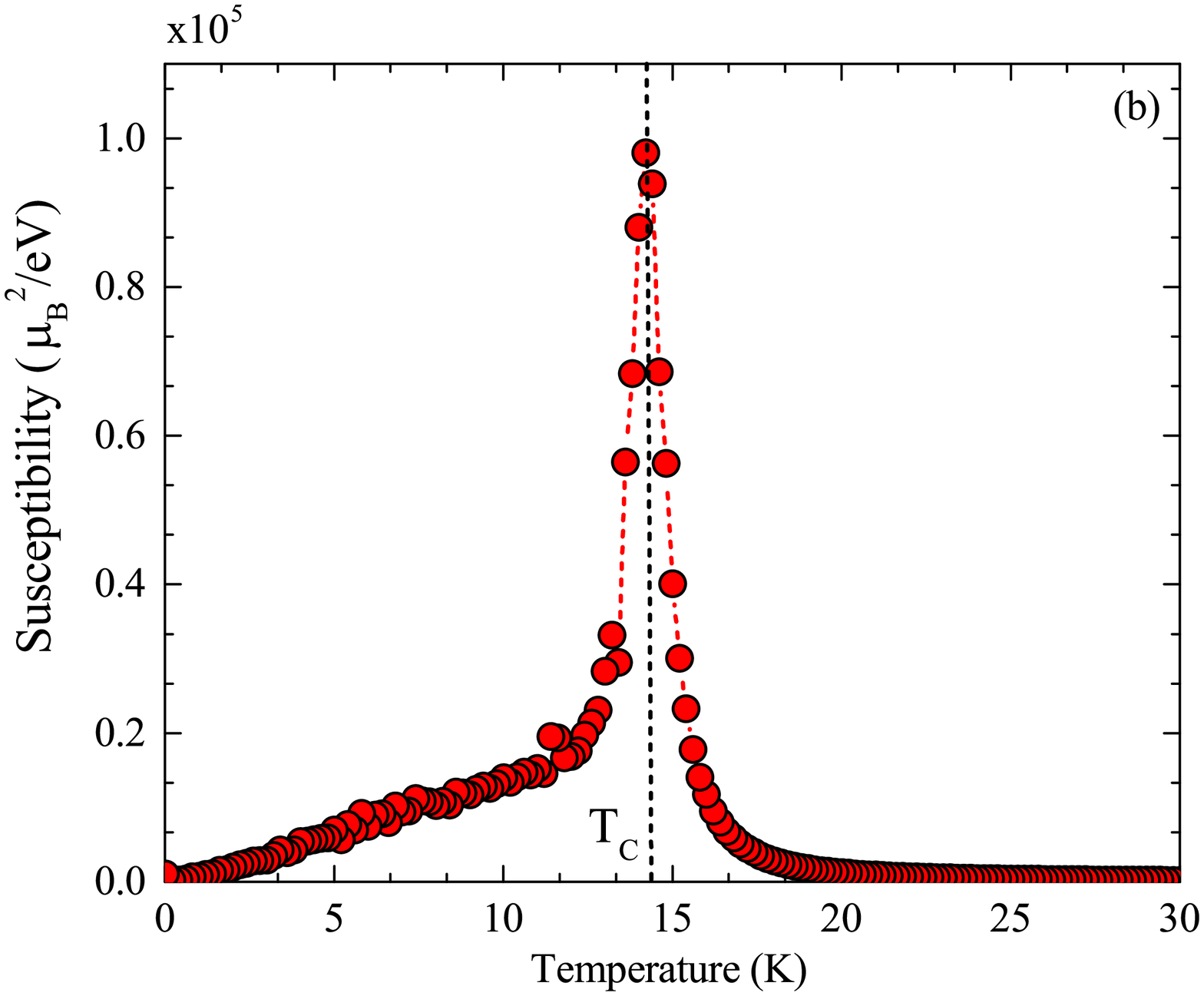}
\includegraphics[scale=0.3]{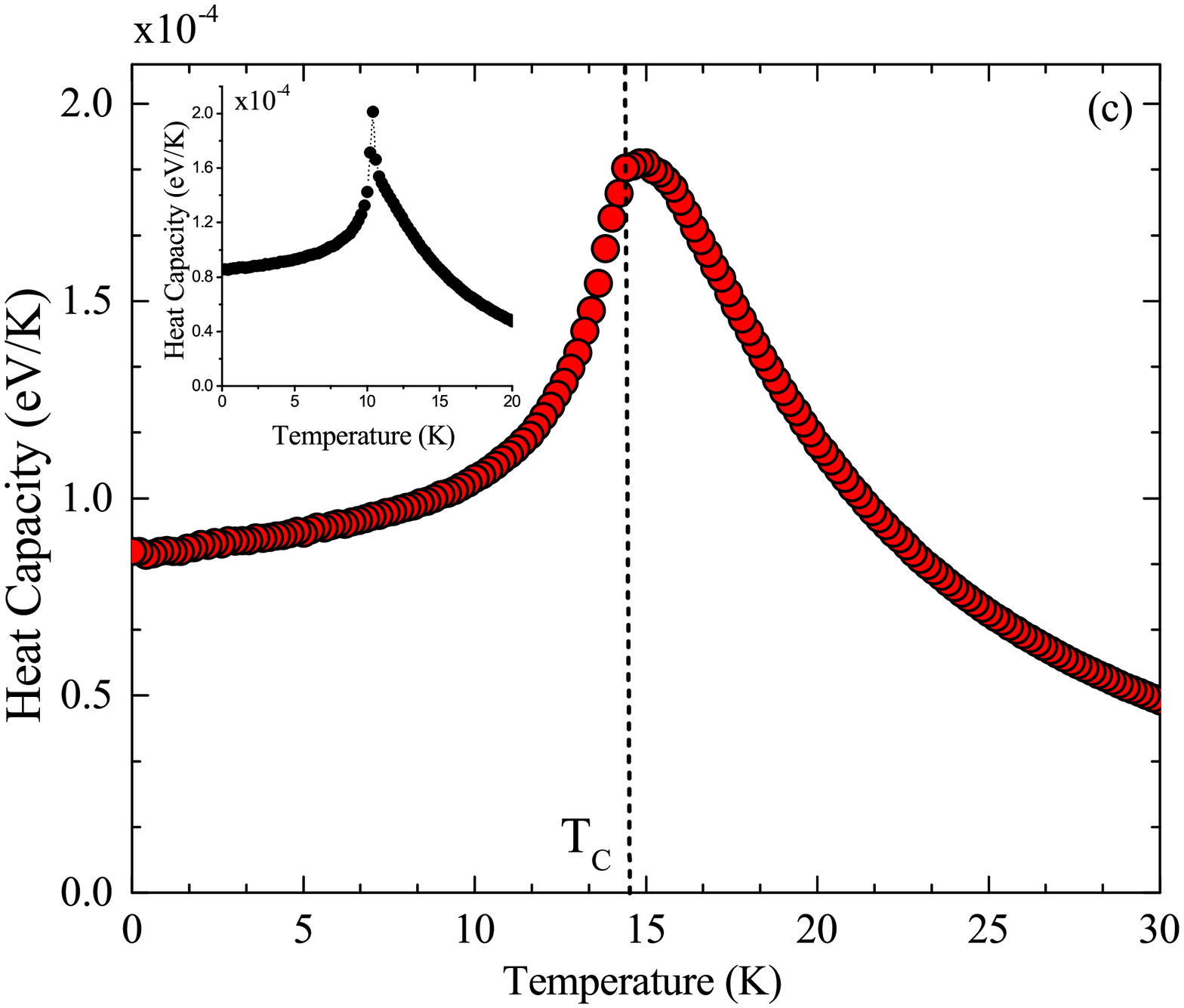}
\caption{(Color online) (a) Variations of the magnetic moment per site, (b) magnetic susceptibility
and (c) specific heat with respect to the temperature. For the sake of completeness, we illustrate the specific heat curve as a function of
the temperature of the bulk system in the inset of Fig. \ref{fig6} (c)}
\label{fig6}
\end{figure}

One can clearly elucidate the magnetic properties of the $\alpha$-RuCl$_3$ monolayer using all spin-spin coupling terms and magnetic anisotropy energies, namely $J_{1}$, $J_{2}$, $J_{3}$, $k_{x}$ and $k_{y}$, which are defined in Section~\ref{comp}. Thermal variations of magnetization, magnetic susceptibility and specific heat curves are illustrated in Figure \ref{fig6}. Our simulation results suggest that the Curie temperature $T_{C}$ is $14.21\ K$. Below this $T_{C}$, the $\alpha$-RuCl$_3$ monolayer exhibits ferromagnetic character. As shown in Figure~\ref{fig6}(a), when the temperature is increased starting from relatively lower temperature value, the magnetization as a function of temperature begins to decrease due to the thermal agitations. Finally,  it disappears when the temperature reaches the critical temperature value, and in other words, the system shows a magnetic phase transition between ordered and disordered phases. As displayed in Figure~\ref{fig6}(b-c), magnetic susceptibility and specific heat curves display a behavior which tend to diverge as the temperature reaches the Curie temperature. For the sake of completeness, we calculate the specific heat as a function the temperature for the bulk system, as shown in the inset of Fig. \ref{fig6}(c). Spin-spin couplings of the bulk $\alpha$-RuCl$_{3}$ noted in section~\ref{comp} are used for this calculation.
Note that we followed the same simulation protocol defined for the monolayer
$\alpha$-RuCl$_3$ because magnetic interlayers are only weakly bonded with the van der Waals
force  ~\cite{plumb2014alpha, sears2015magnetic, banerjee2016proximate} .
We know today from the previously published studies ~\cite{Fletcher1}  that $\alpha-$RuCl$_{3}$ is antiferromagnetic
at the relatively lower temperature regions. Our Monte Carlo simulations support this
fact.  It is clear from  the figure that a phase transition takes place between antiferromagnetic and paramagnetic phases as the temperature reaches the N\'eel temperature. Our MC simulation results indicate the N\'eel temperature as $T_{N}=10.21K$, which is in accordance with the previously obtained results ranging from 6.5K and 15.6K. ~\cite{sears2015magnetic, johnson2015monoclinic}.

\section{Conclusions}\label{conc}
In summary, we used DFT calculations to investigate the stability of RuCl$_3$ and to determine the electronic and magnetic properties of it. The CE values are smaller than of graphite CE, which means the $\alpha$-RuCl$_3$ monolayer can be easily obtained from its bulk phase. According to our DFT calculations, each ruthenium atom gives 1$\mu_B$ magnetic moment to the system, so we obtained 4$\mu_B$ total magnetic moment for per square cell. Monolayer RuCl$_3$ has 3 meV band gap around $\Gamma-M$ high symmetry points and this gap increases to 57 meV by adding SOC effect for spin-down channels, while this gap becomes 1.93 eV for spin-up channel. Furthermore, we implemented Monte Carlo simulation based on Heisenberg model to determine  temperature dependencies of the magnetic properties of the monolayer $\alpha$-RuCl$_3$ system.  According to the our simulation findings, present system  demonstrate a magnetic phase transition between ordered and disordered phases at the Curie temperature 14.21 K. In addition, we performed an additional Monte Carlo simulation using the same protocol defined for the monolayer  to have a better understanding of the behavior the bulk $\alpha$-RuCl$_3$ system with respect to the varying temperature. We find that the system shows a phase transition between antiferromagnetic and paramagnetic phases, with the estimated N\'eel temperature as $10.21$ K. We believe that the results obtained in this study would  be  beneficial for the future theoretical and experimental research on the monolayer $\alpha$-RuCl$_3$ system.

\section{Acknowledgments}
his work was supported by the Scientific and Technological Research Council of Turkey (TUBITAK) under the Research Project No. 117F133. Computing resources used in this work were provided by the TUBITAK ULAKBIM, High Performance and Grid Computing Center (Tr-Grid e-Infrastructure).


\bibliography{rsc} 
\bibliographystyle{rsc} 

\newpage

\providecommand{\noopsort}[1]{}\providecommand{\singleletter}[1]{#1}%

\end{document}